\documentclass[a4paper]{jpconf}
\usepackage{graphicx}
\begin{document}
\title{Experimental study of $\alpha$-induced reactions on $^{64}$Zn for the astrophysical $\gamma$-process}

\author{Gy Gy\"urky$^1$, J Farkas$^1$, Z Hal\'asz$^1$, Zs F\"ul\"op$^1$, E Somorjai$^1$, T Sz\"ucs$^1$, P Mohr$^{1,2}$, A Wallner$^{3,4}$}

\address{$^1$ Institute of Nuclear Research (ATOMKI), H-4001 Debrecen, POB.51., Hungary}
\address{$^2$ Diakonie-Klinikum, D-74523 Schw\"abisch Hall, Germany}
\address{$^3$ Faculty of Physics, University of Vienna, 1090 Wien, Austria}
\address{$^4$ ANSTO, Locked Bag 2001, Kirrawee DC NSW 2232, Australia}
\ead{gyurky@atomki.hu (Gy.\,Gy\"urky)}

\begin{abstract}

For the synthesis of the heavy, proton rich isotopes in the astrophysical $\gamma$-process the precise knowledge of $\alpha$-induced cross sections is of high importance. We have initiated a comprehensive study of the $^{64}$Zn\,+\,$\alpha$ system involving the cross section measurement of different reaction channels as well as the elastic scattering at low, astrophysically relevant energies. In this paper the experimental technique and some preliminary results of the $^{64}$Zn($\alpha$,p)$^{67}$Ga cross section measurement are presented.

\end{abstract}

\section{Introduction}

In various processes of heavy element nucleosynthesis, like e.g.~in the astrophysical $\gamma$-process \cite{arn03}, the reaction rates are calculated from theoretical cross sections since experimental data are scarce. In this mass region the theoretical cross sections are typically obtained from the Hauser--Feshbach statistical model. In this model the cross section can be described by a formula similar to the Breit--Wigner formula for isolated resonances:

\begin{equation}
	\sigma \sim \frac{\Gamma_{\mathrm{in}}\Gamma_{\mathrm{out}}}{\Gamma_{\mathrm{tot}}}
\end{equation}
where $\Gamma_{\mathrm{in}}$, $\Gamma_{\mathrm{out}}$ and $\Gamma_{\mathrm{tot}}$ are the widths of the entrance and the exit channels and the total, respectively. In the statistical model, transmission coefficients $T_i$ take the role of the widths $\Gamma_i$ of the Breit-Wigner formalism. If an $\alpha$-induced reaction is considered, at low, astrophysically relevant energies $\Gamma_{\mathrm{in}}$ is determined mainly by the Coulomb barrier penetration and can be described by a properly chosen $\alpha$-nucleus optical potential. Low energy optical potentials show a significant ambiguity, different global optical potentials lead to largely different reaction cross sections when they are used in statistical models (see e.g.~\cite{gyu10}). Therefore, the study of the $\alpha$-nucleus optical potential deserves a special attention.

The optical potential can be studied experimentally by carrying out high precision elastic scattering experiment over a broad angular range \cite{kis09}. The deviation from the Rutherford cross section can be compared with the predictions of global optical potentials and local potentials can also be constructed. On the other hand, if the elastic scattering cross section measurement is carried out in the nearly full angular range, it can give direct information on the total cross section since the flux which is missing from the elastic channel can be attributed to the sum of all non-elastic channels \cite{moh10}. 

If the cross section of different reaction channels can be measured along with the elastic scattering at the same energies, a more complete description of a given reaction can be achieved. Therefore a research program has been started for the comprehensive study of the $^{64}$Zn\,+\,$\alpha$ system.  

\section{Alpha-induced reactions on $^{64}$Zn}

\begin{figure}
\begin{center}
\includegraphics[angle=-90,width=0.6\textwidth]{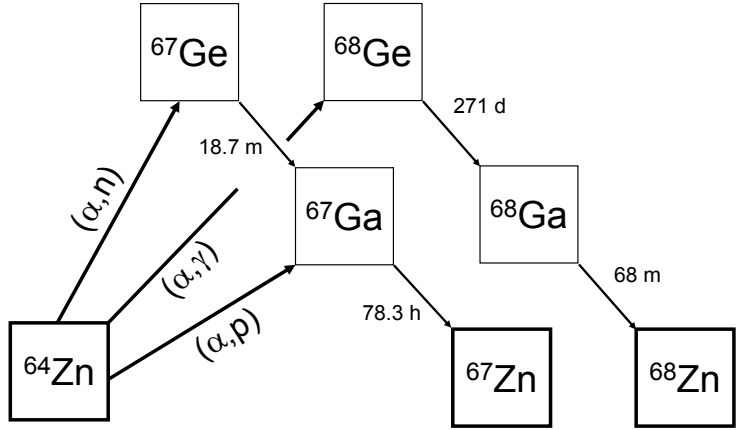}
\caption{\label{fig:reactions} Alpha-induced reactions on $^{64}$Zn. The half-lives of the decays of the reaction products are also indicated.}
\end{center}
\end{figure}

At a typical $\gamma$-process temperature of 3\,GK the Gamow window of the $^{64}$Zn\,+\,$\alpha$ reaction is in between 4 and 6.5\,MeV. In this energy range only the $^{64}$Zn($\alpha,\gamma$)$^{68}$Ge and $^{64}$Zn($\alpha$,p)$^{67}$Ga reaction channels are open (the first one is exothermic while the latter has a threshold at 4.2 MeV). In order to have substantial deviation from the Rutherford cross section, the elastic scattering, however, may be measured only at higher energies, where the $^{64}$Zn($\alpha$,n)$^{67}$Ge channel is also open. Figure \ref{fig:reactions} shows schematically these reactions. In all three cases the reaction products are radioactive, thus the cross section can be measured with the activation method. The aim of the present work is to measure all three channels as well as the elastic scattering\footnote{Some other channels with multiple particle emission are also open, but the cross sections of these reactions are typically much weaker than that of the studied ones.}. The first step is the measurement of the $^{64}$Zn($\alpha$,p)$^{67}$Ga reaction. Since the product of the $^{64}$Zn($\alpha$,n)$^{67}$Ge reaction decays with a short half-life to the product of the $^{64}$Zn($\alpha$,p)$^{67}$Ga reaction, and the separation of the two reactions by the decay counting leads to increased uncertainty, the $^{64}$Zn($\alpha$,p)$^{67}$Ga cross section has been measured first below the ($\alpha$,n) threshold located at E$_\alpha$\,=\,9.56\,MeV. 

\section{Experimental procedure}

\begin{figure}
\begin{center}
\includegraphics[angle=-90,width=0.8\textwidth]{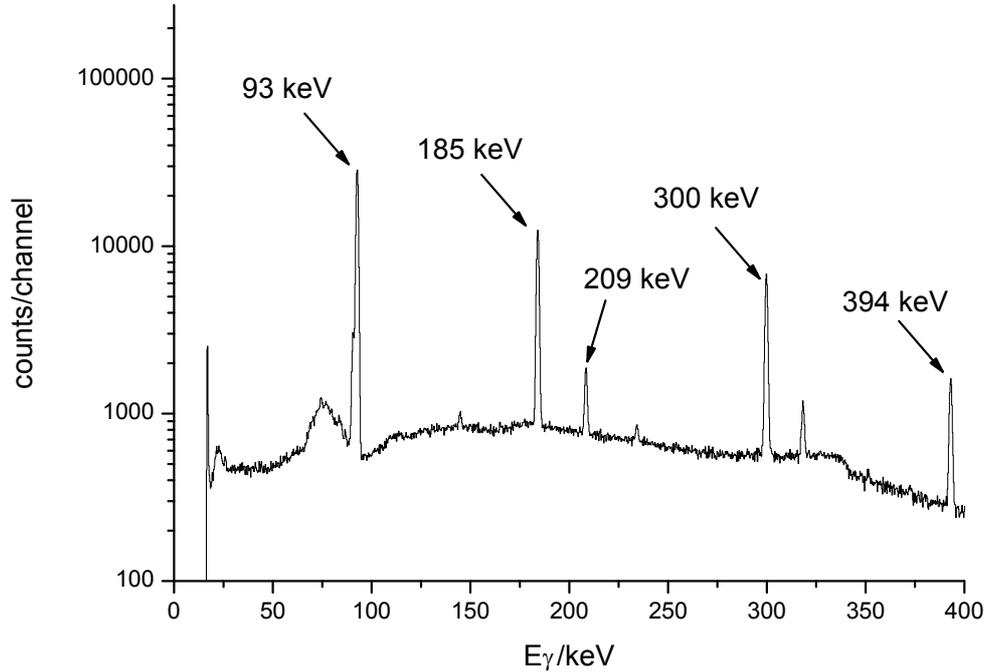}
\caption{\label{fig:spectrum} Typical $\gamma$-spectrum taken for 14 hours on the target irradiated with 8.6 MeV $\alpha$-beam. The peaks used for the analysis are indicated by arrows.}
\end{center}
\end{figure}

Natural isotopic composition zinc targets have been prepared by evaporating metallic Zn onto thin Al foil backings. The target thickness has been measured by weighing and in some cases checked by Rutherford Backscattering (RBS). Further RBS measurements are in progress for the target characterization. The targets have been irradiated by the $\alpha$-beam of the cyclotron accelerator of ATOMKI. The beam current was kept below 1\,$\mu$A in order not to have target deterioration (the target stability was continuously monitored by RBS). In the first experimental campaign the cross section has been measured at nine different $\alpha$-energies between 6.3 and 9.5\,MeV. The irradiations lasted between 5 and 18 hours depending on the energy. The induced activities have been measured by detecting the $\gamma$-rays following the electron capture decay of the $^{67}$Ga reaction product with a 100\,\% relative efficiency HPGe detector equipped with 4\,$\pi$ shielding. The decay of $^{67}$Ga was followed for at least 12 hours by detecting its strongest $\gamma$-lines at 93.3, 184.6, 209.0, 300.2 and 393.5 keV. The absolute efficiency of the detector has been measured with several calibrated sources. Figure \ref{fig:spectrum} shows a typical $\gamma$-spectrum taken on the target irradiated with 8.6\,MeV $\alpha$-beam. The peaks used for the analysis are indicated by arrows.

\section{Results and outlook}

\begin{figure}
\begin{center}
\includegraphics[angle=-90,width=0.8\textwidth]{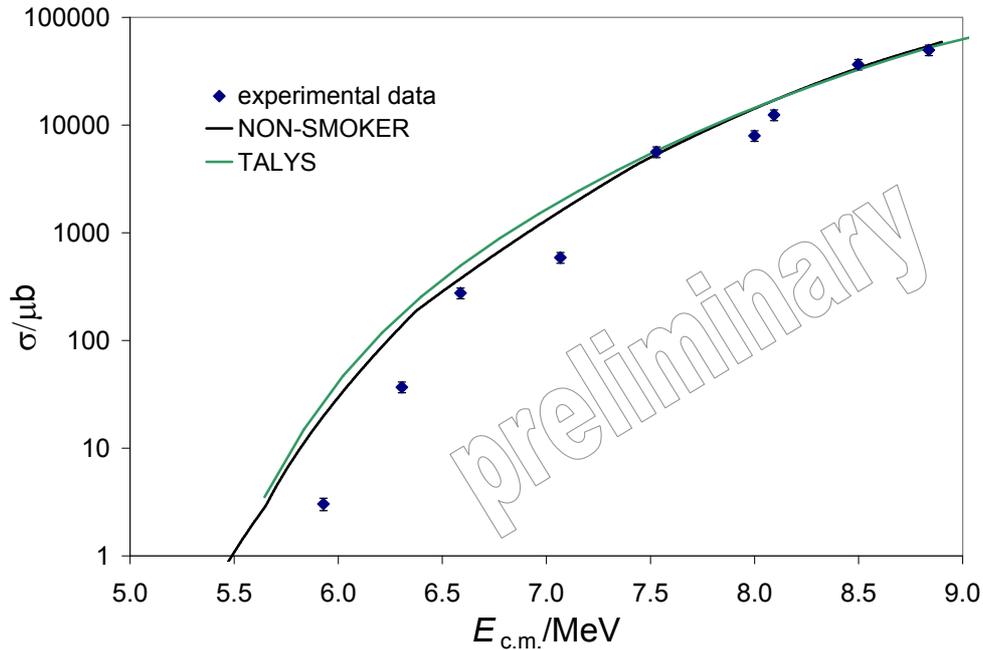}
\caption{\label{fig:results} Preliminary results of the $^{64}$Zn($\alpha$,p)$^{67}$Ga cross section measurement. The experimental data as well as predictions obtained with the NON-SMOKER and TALYS codes are shown.}
\end{center}
\end{figure}

The $^{64}$Zn($\alpha$,p)$^{67}$Ga cross section has been measured below the ($\alpha$,n) threshold in the center-of-mass energy range between 5.9 and 8.8 MeV which is at the upper edge and above the relevant Gamow window. In this energy range the cross section varies between 3 and 5$\cdot$10$^4$\,$\mu$b. The preliminary results can be seen in Figure \ref{fig:results} where the cross sections are plotted versus the c.m. energy. Also shown are the calculations carried out by the NON-SMOKER \cite{NONSMOKER} and TALYS \cite{TALYS} codes. In general the agreement between experiment and theory is reasonably good, although stronger deviations can be found towards lower energies. Further measurements, the final analysis of the data and the comparison with theoretical predictions are in progress.

In principle, the $^{64}$Zn($\alpha,\gamma$)$^{68}$Ge reaction can also be studied in the investigated energy range by activation. Some experimental difficulties are, however, encountered. The half-life of $^{68}$Ge is relatively long (271 days) making the activity of the irradiated targets rather low. Moreover, the decay of $^{68}$Ge is not followed by any $\gamma$-emission. Its decay product, $^{68}$Ga is also radioactive and emits some $\gamma$-rays, but the relative intensity of its strongest $\gamma$-line is only 3\,\%. Thus the $\gamma$-detection based activation measurement of the $^{64}$Zn($\alpha,\gamma$)$^{68}$Ge is challenging. Some attempts have been made to observe the $^{68}$Ga decay, but only very low statistics could be obtained in several weeks of counting. To circumvent this problem a different method is planned. $^{68}$Ge has long enough half-life to be measured with Accelerator Mass Spectrometry (AMS) technique. Some development for such an experiment is in progress at the VERA facility \cite{VERA} in Vienna, Austria.

\section*{Acknowledgments}
This work was supported by the European Research Council StG. 203175 and OTKA grants K68801 and NN83261(EuroGENESIS).


\section*{References}
\begin{thebibliography}{9} 
\bibitem{arn03} Arnould M and Goriely S 2003 \textit{Phys. Rep.} \textbf{384} 1
\bibitem{gyu10} Gy\"urky Gy \textit{et al.} 2010 \textit{J. Phys. G} \textbf{37} 115201
\bibitem{kis09} Kiss G G \textit{et al.} 2009 \textit{Phys. Rev. C} \textbf{80} 045807
\bibitem{moh10} Mohr P \textit{et al.} 2010 \textit{Phys. Rev. C} \textbf{82} 047601
\bibitem{NONSMOKER} Rauscher T and Thielemann F-K 2001 \textit{At. Data Nucl. Data Tables} {\bf 79} 47, www.nucastro.org
\bibitem{TALYS} Koning A J, Hilaire S and Duijvestijn M C 2008 \textit{Proceedings of the International Conference on Nuclear Data for Science and Technology - ND2007} 211
\bibitem{VERA} Wallner A 2010 \textit{Nucl. Instr. and Meth.} \textbf{B268} 1277
\end{thebibliography}
\end{document}